\documentclass[dvips,12pt]{article}
\usepackage{graphicx,color}
\newcommand\figcaption{\def\@captype{figure}\caption}
\newcommand\tabcaption{\def\@captype{table}\caption}
\def\al{\alpha}
\def\kp{\kappa}
\def\nb{\nabla}
\def\pa{\partial}

\def\Om{\Omega}
\def\om{\omega}
\def\Ga{\Gamma}
\def\ga{\gamma}
\def\be{\beta}

\def\dl{\delta}

\def\La{\Lambda}
\def\lan{\langle}
\def\ran{\rangle}

\def\sg{\sigma}
\def\wt{\widetilde}

\def\diag{\mbox {diag}}

\title{\LARGE\bf The covariant derivatives and energy momentum tensor of spinors}

\author{Ying-Qiu Gu\footnote {email: yqgu@luody.com.cn},
\and Bijan Saha\footnote {email: bijan@jinr.ru}}
\date {\footnotesize  Department of Mathematics, Fudan University, Shanghai, 200433, China\\
 \vspace*{2.0ex}
 Laboratory of Information Technologies\\ Joint Institute for
Nuclear Research 141980 Dubna, Russia}

\begin{document}
\maketitle\DeclareGraphicsRule{.eps.gz}{eps}{.eps.bb}{`gunzip -c
#1}

\begin{abstract}
To understand the coupling behavior of the spinor with spacetime,
the explicit form of the energy-momentum tensor of the spinor in
curved spacetime is important. This problem seems to be overlooked
for a long time. In this paper we derive the explicit form of
energy momentum tensors and display some equivalent but simple
forms of the covariant derivatives for both the Weyl spinor and
the Dirac bispinor in curved spacetime.
\end{abstract}

\section{Introduction}
\setcounter{equation}{0}

The spinor field is the most sensitive to the gravitational one.
This very property of the spinor field leads many researchers to
investigate the self-consistent system of the spinor and the
gravitational fields in recent years. These works gave rise to a
number of interesting results. The introduction of the spinor
field into the system in question results in an accelerated mode
of evolution \cite{1,2}, while some special choice of spinor field
nonlinearity provides singularity-free cosmological solutions
\cite{3,4}. One of the authors[Y.Q. Gu]  once computed the
numerical solution of the spinor soliton, and found it has
discrete spectrum \cite{gu}. In these works, the authors mainly
computed the results in the diagonal metric, and in this case the
energy momentum tensor can be directly derived from Lagrangian of
the spinor field. But for general case, the energy momentum tensor
quoted seems incomplete, and this problem seems to be overlooked.
In what follows we give a systematic review of the problem and
derive some useful and convenient results.

\section{Simplification of spinor Covariant derivatives}
\setcounter{equation}{0}

The covariant derivatives of a spinor have been constructed by
several authors \cite{6,7,8,9,10}, but their formalisms are somewhat different
in form due to the not unique representation of the products of
the Pauli and Dirac matrices. In this paper, we choose the Pauli
and Dirac matrices in flat spacetime as follows:
\begin{equation}
 \sg^{\mu}\equiv \left \{\pmatrix{
 1 & 0 \cr 0 & 1},\pmatrix{
 0 & 1 \cr 1 & 0},\pmatrix{
 0 & -i \cr i & 0},\pmatrix{
 1 & 0 \cr 0 & -1}
 \right\},\label{1.1}\end{equation}
\begin{equation}
\wt{\sg}^{\mu}\equiv(\sg^0,-\vec\sg),\qquad \vec\sg=(\sg^1,\sg^2,\sg^3).
\label{1.2} \end{equation}
\begin{equation}
\begin{array}{l}
\ga^\mu\equiv \pmatrix{0 & \sg^\mu\cr\wt\sg^\mu & 0}
=\left\{\pmatrix{0 & I\cr I & 0},\pmatrix{ 0 &\vec\sg\cr
-\vec\sg&0}\right\}.\end{array} \label{1.3} \end{equation}
In the flat
spacetime, the Dirac equation for bispinor $\phi$ has the form
\begin{equation}
\ga^\mu i\pa_\mu\phi=m\phi. \label{1.4}
 \end{equation}
Introducing
\begin{equation}
\phi=\pmatrix{ \wt\psi\cr \psi}, \label{1.5} \end{equation}
where
$\psi,~\wt\psi$ are two Weyl spinors, Eq. (\ref{1.4}) can be written as
\begin{equation}\left\{\begin{array}{l}
\sg^\mu i\pa_\mu\psi=m\wt\psi,\\
\wt\sg^\mu i\pa_\mu\wt\psi=m\psi.
\end{array}\right. \label{1.6}\end{equation}
Eq. (\ref{1.6}) is the so-called chiral representation in gauge theory.

Assume $x^\mu$ be the coordinates, $d \bar x^a$ be the local frame, then
using the vierbein formalism for the curved spacetime we can write
\begin{equation}\begin{array}{l}
d x^\mu=h^\mu_{~a} d\bar x^a,\quad d x_\mu=l_\mu^{~a} d\bar
x_a,\quad  h^\mu_{~a} l_\mu^{~b}=h^\mu_{~a} l_\mu^{~b}=\dl^b_a.
\end{array}\label{1.7}\end{equation}
or equivalently
\begin{equation}
g^{\mu\nu}=h^\mu_{~a} h^\nu_{~b} \eta^{ab},\qquad
g_{\mu\nu}=l_\mu^{~a} l_\nu^{~b} \eta_{ab}, \label{1.8}
\end{equation} where $\eta_{ab}=\diag[1,-1,-1,-1]$ is the
Minkowski metric.

Denote the Pauli matrices in curved spacetime by
\begin{equation}\left\{\begin{array}{l}
\rho^\mu=h^\mu_{~a}\sg^a,\qquad\rho_\mu=l_\mu^{~a}\sg_a, \\
\wt\rho^\mu=h^\mu_{~a}\wt\sg^a,\qquad
\wt\rho_\mu=l_\mu^{~a}\wt\sg_a,
\end{array}\right.\label{1.9}\end{equation}
then we have
\begin{equation}\rho^\mu\wt\rho^\nu+\rho^\nu\wt\rho^\mu=\wt\rho^\mu\rho^\nu+\wt\rho^\nu\rho^\mu=2
g^{\mu\nu}. \label{1.10}\end{equation} Here we start with the
formalism given by Peter G. Bergmann \cite{8}. The field equation
(\ref{1.6}) in curved spacetime is given by
\begin{equation}\left\{\begin{array}{l}
\rho^\mu i\nb_\mu\psi=m\wt\psi,\\
\wt\rho^\mu i\wt\nb_\mu\wt\psi=m\psi,
\end{array}\right. \label{1.11}\end{equation}
where $\nb_\mu=\pa_\mu+\Ga_\mu$,$~\wt\nb_\mu=\pa_\mu+\wt\Ga_\mu$
are the covariant derivatives acting on $\psi$ and $\wt\psi$, respectively. $\Ga_\mu$
and $\wt\Ga_\mu$ are the spinor affine connections satisfying (see
Appendix A)
\begin{equation}\Ga_\mu=\frac 1 4 \wt\rho_\al\rho^\al_{;\mu},\qquad
\wt\Ga_\mu=\frac 1 4 \rho_\al\wt\rho^\al_{;\mu}, \label{1.12}\end{equation}
where $\rho^\al_{;\mu}=\pa_\mu\rho^\al+\Ga^\al_{\mu\be}\rho^\be$
and
$\wt\rho^\al_{;\mu}=\pa_\mu\wt\rho^\al+\Ga^\al_{\mu\be}\wt\rho^\be$.

By the representation of vierbein\cite{gu1}, we have the following
expression for $\Ga_\mu$:
\begin{eqnarray}
\begin{array}{lll}
\Ga_\mu &=& \frac 1 4 \wt\rho_\al(\pa_\mu\rho^\al+\Ga^\al_{\mu\nu}\rho^\nu)\\
&=&\frac 1 4 \wt\rho_\al[{\pa_\mu\rho^\al}+\frac 1 2
g^{\al\be}(\pa_\mu g_{\be\nu}+\pa_\nu
g_{\be\mu}-\pa_\be g_{\mu\nu})\rho^\nu]\\
&=& \frac 1 4 \wt\rho_\al\pa_\mu\rho^\al+\frac 1 {16} \pa_\mu
g_{\be\nu}(\wt\rho^\be\rho^\nu+\wt\rho^\nu\rho^\be)-\frac 1 8
\pa_\al
g_{\mu\be} (\wt\rho^\al\rho^\be-\wt\rho^\be\rho^\al)\\
 &=&\frac 1 4 \wt\rho_\al\pa_\mu\rho^\al+\frac 1 8 \pa_\mu
g_{\be\nu}g^{\be\nu}-\frac 1 8 \pa_\al g_{\mu\be}
(\wt\rho^\al\rho^\be-\wt\rho^\be\rho^\al)\\
 &=&\frac 1 4 (\wt\rho^\al\pa_\mu\rho_\al-\Ga_{\mu\nu}
^{\nu})-\frac 1 8 \pa_\al g_{\mu\be}
(\wt\rho^\al\rho^\be-\wt\rho^\be\rho^\al).
\end{array} \label{1.14}\end{eqnarray}
Similarly we have
\begin{eqnarray}
\wt\Ga_\mu=\frac 1
4(\rho^\al\pa_\mu\wt\rho_\al-\Ga_{\mu\nu}^\nu)-\frac 1 8 \pa_\al
g_{\mu\be} (\rho^\al\wt\rho^\be-\rho^\be\wt\rho^\al). \nonumber
\end{eqnarray}
By (\ref{1.14}) and the above expression, we find
$\mbox{tr}(\Ga_\mu)=\mbox{tr}(\wt\Ga_\mu)=0$, this means that
$\Ga_\mu$ and $\wt\Ga_\mu$ have just geometrical effects(see
Appendix A).

Defining
\begin{equation}
\varrho^\mu \equiv
h^\mu_\nu\ga^\nu,~~{\mbox{with}}~~\varrho^\mu\varrho^\nu+\varrho^\nu\varrho^\mu=2
g^{\mu\nu}.\label{1.15}\end{equation} For Dirac bispinor $\phi$,
we have the following covariant derivative
\begin{equation}
\nb_\mu\phi=(\pa_\mu+\bar\Ga_\mu)\phi,\qquad \bar\Ga_\mu=\frac 1 4
\varrho_\nu\varrho^\nu_{;\mu}, \label{1.16} \end{equation} with
$\bar\Ga_\mu$ being the Fock-Ivannenko connection.

Analogous to (\ref{1.14}) for $\bar\Ga_\mu$ one finds
\begin{equation}
\bar\Ga_\mu=\frac 1 4
(\varrho^\al\pa_\mu\varrho_\al-\Ga_{\mu\al}^\al)-\frac 1 8 \pa_\al
g_{\mu\be}(\varrho^\al\varrho^\be-\varrho^\be\varrho^\al).\label{1.17}\end{equation}
The expressions (\ref{1.14}) and (\ref{1.17}) are more convenient
for calculation and analyzing than (\ref{1.12}) and (\ref{1.16}).
In what follows, we will see the convenience.

\section{Energy momentum tensor of the spinors}
\setcounter{equation}{0}

Let us go back to the Eq. (\ref{1.11}). Define the total connection of the spinors as
\begin{eqnarray}\left\{\begin{array}{l}
\Ga\equiv \rho^\mu\Ga_\mu= \frac 1 4\rho^\mu
(\wt\rho^\al\pa_\mu\rho_\al-\Ga_{\mu\al}^\al)-\frac 1 8 \pa_\al
g_{\mu\be}\rho^\mu
(\wt\rho^\al\rho^\be-\wt\rho^\be\rho^\al),\\
\wt\Ga\equiv \wt\rho^\mu\wt\Ga_\mu=\frac 1 4\wt\rho^\mu
(\rho^\al\pa_\mu\wt\rho_\al-\Ga_{\mu\al}^\al)-\frac 1 8 \pa_\al
g_{\mu\be}\wt\rho^\mu
(\rho^\al\wt\rho^\be-\rho^\be\wt\rho^\al).\end{array}\right.
\label{2.1}\end{eqnarray} Considering that all
$\rho^\mu,~\wt\rho^\mu$ are Hermitian and $g_{\mu\nu}=g_{\nu\mu}$,
by (\ref{1.10}) we have
\begin{eqnarray}
\begin{array}{lll}
\Ga^+ &=&  \frac 1 4(\rho^\mu \wt\rho^\al\pa_\mu\rho_\al)^+-\frac
1 4\rho^\mu\Ga_{\mu\al}^\al-\frac 1 8 \pa_\al
g_{\mu\be} (\rho^\be\wt\rho^\al\rho^\mu-\rho^\al\wt\rho^\be\rho^\mu) \\
&=&\frac 1 4\pa_\mu\rho_\al\wt\rho^\al\rho^\mu-\frac 1
4\rho^\mu\Ga_{\mu\al}^\al-\frac 1 8 \pa_\al
g_{\mu\be} [\rho^\mu\wt\rho^\al\rho^\be-\rho^\al\frac 1 2(\wt\rho^\be\rho^\mu+\wt\rho^\mu\rho^\be)]  \\
&=&\frac 1 4\pa_\mu\rho_\al\wt\rho^\al\rho^\mu-\frac 1
4\rho^\mu\Ga_{\mu\al}^\al-\frac 1 8 \pa_\al g_{\mu\be}
(\rho^\mu\wt\rho^\al\rho^\be-g^{\mu\be}\rho^\al),
\end{array}
\end{eqnarray}
so we have
\begin{equation}
\Om\equiv\frac i 2(\Ga-\Ga^+) =  \frac i 8(\rho^\mu
\wt\rho^\al\pa_\mu\rho_\al-\pa_\mu\rho_\al\wt\rho^\al\rho^\mu).
\end{equation}
Similarly
\begin{equation}
\wt\Om\equiv\frac i 2(\wt\Ga-\wt\Ga^+) =  \frac i 8(\wt\rho^\mu
\rho^\al\pa_\mu\wt\rho_\al-\pa_\mu\wt\rho_\al\rho^\al\wt\rho^\mu).
\end{equation}
For diagonal metric, we have $\Om=\wt\Om=0$\cite{gu1}.

Now we consider the interacting system of spinor and gravitational
fields. The spinor field is chosen to be the one satisfying Eq.
(\ref{1.11}) while the gravitational field is given by some metric
functions $g_{\mu\nu}$. Given $g_{\mu\nu}$ the Ricci tensor and
scalar curvature can be defined from
\begin{eqnarray}
R_{\mu\nu}=\pa_\mu\Ga^\al_{\nu\al}-\pa_\al\Ga^\al_{\mu\nu}+\Ga^\al_{\mu\be}\Ga^\be_{\nu\al}-
\Ga^\al_{\mu\nu}\Ga^\be_{\al\be},~~ R\equiv g^{\mu\nu}R_{\mu\nu}. \nonumber
\end{eqnarray}
The total Lagrangian of the interacting system reads
\begin{equation}
{\cal L}=\frac 1 {2\kp}{\cal L}_g+{\cal L}_m=-\frac1
{2\kp}(R+2\Lambda)+{\cal L}_m, \label{lag}
\end{equation}
where $\kp=8\pi G$, $\Lambda$ stands for the cosmological factor,
${\cal L}_m$ is the Lagrangian of spinors. Noticing $\rho^\mu$ and
$\wt\rho^\mu$ are all Hermitian, we have
\begin{eqnarray} \begin{array}{lll}
{\cal L}_m &=& {\rm Re}\lan\psi^+\rho^\mu
i\nb_\mu\psi+\wt\psi^+\wt\rho^\mu
i\wt\nb_\mu\wt\psi\ran-m(\wt\psi^+\psi+\psi^+\wt\psi),\\
&=&\frac 1 2 [\psi^+\rho^\mu i\pa_\mu\psi+(i\pa_\mu\psi)^+\rho^\mu
\psi+\wt\psi^+\wt\rho^\mu
i\pa_\mu\wt\psi+(i\pa_\mu\wt\psi)^+\wt\rho^\mu \wt\psi]+\\
&~&\frac
i 2 [\psi^+(\Ga-\Ga^+)\psi+\wt\psi^+(\wt\Ga-\wt\Ga^+)\wt\psi]-m(\wt\psi^+\psi+\psi^+\wt\psi),\\
&=&{\rm Re} \lan\psi^+\rho^\mu i\pa_\mu\psi+\wt\psi^+\wt\rho^\mu
i\pa_\mu\wt\psi
\ran+\psi^+\Om\psi+\wt\psi^+\wt\Om\wt\psi-m(\wt\psi^+\psi+\psi^+\wt\psi).
\end{array}
\label{2.2} \end{eqnarray} From (\ref{2.2}) we learn that, the
total connection $\Ga$ and $\wt\Ga$ will vanish in ${\cal L}_m$
for diagonal metric. Variation of the Lagrangian (\ref{lag}) with
respect to $g_{\mu\nu}$ gives the the Einstein's equation
\begin{equation}
R^{\mu\nu}-\frac 1 2g^{\mu\nu}R-\La g^{\mu\nu}=\kp T^{\mu\nu},
\end{equation}
where $T^{\mu\nu}$ is the energy momentum tensor of the material field which is defined by
\begin{equation}
T^{\mu\nu}=-2\frac {\dl({\cal L}_m \sqrt{g})}{\sqrt{g}\dl
g_{\mu\nu}}=-2\frac {\dl{\cal L}_m }{\dl l_\al^{~n}}\frac {\pa
l_\al^{~n}}{\pa g_{\mu\nu}}-g^{\mu\nu}{\cal L}_m ,\label{2.3}
\end{equation}
where $g=|\det(g_{\mu\nu})|$ and $\frac \dl {\dl l_\al^{~n}}$ is
the Euler derivative defined by
\begin{equation}\begin{array}{lll}
\frac {\dl{\cal L}_m }{\dl l_\al^{~n}}&=&\frac {\pa{\cal L}_m
}{\pa l_\al^{~n}}-\frac 1 {\sqrt{g}}\pa_\kp\frac {(\pa{\cal L}_m
\sqrt{g})}{\pa(\pa_\kp l_\al^{~n})}=\frac {\pa{\cal L}_m }{\pa
l_\al^{~n}}-(\pa_\kp +\Ga^\ga_{\kp\ga})\frac {\pa{\cal L}_m
}{\pa(\pa_\kp
l_\al^{~n})}\\
&=& {\mbox{Re}} \lan\psi^+\frac{\pa \rho^\mu}{\pa l_\al^{~n}}
i\pa_\mu\psi+\wt\psi^+\frac{\pa \wt\rho^\mu}{\pa l_\al^{~n}}
i\pa_\mu\wt\psi \ran+\psi^+\frac{\pa \Om}{\pa
l_\al^{~n}}\psi+\wt\psi^+\frac{\pa
\wt\Om}{\pa l_\al^{~n}}\wt\psi \\
&~&-(\pa_\kp +\Ga^\ga_{\kp\ga})\left(\psi^+\frac{\pa \Om}{\pa
(\pa_\kp l_\al^{~n})}\psi+\wt\psi^+\frac{\pa \wt\Om}{\pa(\pa_\kp
l_\al^{~n})}\wt\psi\right).
\end{array}\label{23}
\end{equation}
For the $LU$ decomposition of $g_{\mu\nu}$, we have the following
equation\cite{gu1}
\begin{equation}\left\{\begin{array}{l}
\frac {\pa l_\al^{~n}}{\pa g_{\mu\nu}} =\frac 1 4 (\dl^\mu_\al
h^\nu_{~m}+\dl^\nu_\al h^\mu_{~m})\eta^{mn}+S^{\mu\nu|n}_\al, \\
S^{\mu\nu|n}_\al=\frac 1 4
\sum^4_{a>b=1}(h^\mu_{~a}h^\nu_{~b}+h^\nu_{~a}h^\mu_{~b})(l_\al^{~a}\eta^{nb}-l_\al^{~b}\eta^{an}),\end{array}
\right. \label{vier}\end{equation} where index 4 means $t$ or
index 0.

The terms in (\ref{23}) can be calculated as follows\cite{gu1},
\begin{equation}\left\{\begin{array}{l}
\frac{\pa \rho^\om}{\pa l_\al^{~n}}=- \rho^\al
h^\om_{~n},~~\frac{\pa \rho^\om}{\pa l_\al^{~n}}\frac {\pa
l_\al^{~n}}{\pa g_{\mu\nu}} =-\frac 1 4 (\rho^\mu
g^{\nu\om}+\rho^\nu g^{\mu\om})-
\rho^\al h^\om_{~n}S^{\mu\nu|n}_\al,\\
\frac{\pa\Om}{\pa l_\al^{~n}} = \frac i 8
h^\be_{~n}(\pa_\be\rho_\om\wt\rho^\om\rho^\al-\rho^\al
\wt\rho^\om\pa_\be\rho_\om)+\frac i 8
h^\om_{~n}(\pa_\be\rho_\om\wt\rho^\al\rho^\be-\rho^\be
\wt\rho^\al\pa_\be\rho_\om), \\
\frac{\pa\Om}{\pa (\pa_\be l_\al^{~n})} = \frac i 8
(\rho^\be\wt\rho^\al\sg_n-\sg_n\wt\rho^\al\rho^\be),\\
~~\wt\rho^\al\frac {\pa \rho_\al}{\pa g_{\mu\nu}}=\frac 1 2
g^{\mu\nu}+\frac 1 2
\sum^4_{a>b=1}(h^\mu_{~a}h^\nu_{~b}+h^\nu_{~a}h^\mu_{~b})\wt\sg^a\sg^b,
\label{term}\end{array}\right.
\end{equation}
\begin{equation}\begin{array}{lll}
K^{\mu\nu}(\psi)&\equiv&(\pa_\kp+\Ga^\ga_{\kp\ga})\left(\psi^+\frac{\pa\Om}{\pa
(\pa_\kp l_\al^{~n})}\psi\right)\cdot\frac{\pa l_\al^{~n}}{\pa
g_{\mu\nu}}\\ &=& \frac i 8(\pa_\kp+\Ga^\ga_{\kp\ga})
\left[\psi^+\left(\rho^\kp\wt\rho^\al \frac{\pa\rho_\al}{\pa
g_{\mu\nu}}-\frac{\pa\rho_\al}{\pa
g_{\mu\nu}}\wt\rho^\al\rho^\kp\right)\psi\right]-\\
&~&\frac i 8 \psi^+\left(\rho^\kp\wt\rho^\al
\pa_\kp\frac{\pa\rho_\al}{\pa
g_{\mu\nu}}-\pa_\kp\frac{\pa\rho_\al}{\pa
g_{\mu\nu}}\wt\rho^\al\rho^\kp\right)\psi, \end{array}\label{K}
\end{equation}
In the skew spacetime, i.e., for the intrinsically diagonal
metric, $\Om\neq 0$ results in the gravimagentic effects, which
couples the spinor fields in a quite complicated manner\cite{gu2}.
Similarly to the Ricci tensor, Eq.(\ref{term}) and (\ref{K}) are
difficult to be simplified due to the complicated form of $\Om$
and $S^{\mu\nu|n}_\al$.

However for the intrinsically diagonal metric, the problem is much
simpler. For this case, we have $\Om=\wt\Om=S^{\mu\nu|n}_\al=0$,
then
\begin{eqnarray}
\begin{array}{lll}
T^{\mu\nu}&=&- 2\frac {\pa{\cal L}_m}{\pa g_{\mu\nu}}-g^{\mu\nu}{\cal L}_m,\\
&=&\frac 1 2{\rm Re}\lan\psi^+(\rho^\mu i\pa^\nu+\rho^\nu
i\pa^\mu)\psi +\wt\psi^+(\wt\rho^\mu i\pa^\nu+\wt\rho^\nu
i\pa^\mu)\wt\psi\ran-g^{\mu\nu}{\cal L}_m,
\end{array}\label{2.4}\end{eqnarray}
$T^{\mu\nu}$ is independent of spinor connection. Similarly, for
Dirac bispinor equation (\ref{1.4}), we have
\begin{equation}
{\cal L}_m = {\rm Re}\lan\phi^+\ga^0\varrho^\mu
i\pa_\mu\phi\ran-m\phi^+\ga^0\phi, \label{2.5}\end{equation}
\begin{equation}
T^{\mu\nu}=\frac 1 2 {\rm Re}\lan\phi^+\ga^0(\varrho^\mu
i\pa^\nu+\varrho^\nu i\pa^\mu)\phi\ran-g^{\mu\nu}{\cal L}_m.
\label{2.6}\end{equation}

\section{Conclusions}
\setcounter{equation}{0}

A self-consistent system of spinor and gravitational fields has
been investigated. The explicit form of energy momentum tensors
and some equivalent but simple forms of the covariant derivatives
for both the Weyl spinor and the Dirac bispinor in curved
spacetime have been derived.

\section*{Acknowledgments}
The first author thanks his supervisor Academician Li Ta-tsien for
guidance and encouragement, Prof. Wang Hao for kind help.

\newpage
\section*{Appendix A: the verification of (1.12)}
The covariant derivatives of the spinor must keep the invariant
form under both the coordinates and the tetrad transformation. To
determine the algebraic conditions of the spinor connection and
its geometrical meanings straightway are quite abstract. However
this problem can be simplified by considering the consistency of
the covariant derivatives of vector and spinor, that is, for Weyl
spionr $\psi$,
$$
q^\mu\equiv\psi^+\rho^\mu\psi \eqno(A1)
$$
is a vector. Then on one hand we have the covariant derivative for
vector
$$
q^\mu_{;\nu}= \pa_\nu
q^\mu+\Ga^\mu_{\nu\om}q^\om=(\pa_\nu\psi)^+\rho^\mu\psi+\psi^+\rho^\mu(\pa_\nu\psi)+\psi^+\rho^\mu_{;\nu}\psi;
 \eqno(A2) $$
on other hand we have the covariant derivative for spinor
$$
\begin{array}{lll}
q^\mu_{;\nu}&=&(\nb_\nu\psi)^+\rho^\mu\psi+\psi^+\rho^\mu(\nb_\nu\psi),\\
&=&(\pa_\nu\psi)^+\rho^\mu\psi+\psi^+\rho^\mu(\pa_\nu\psi)+\psi^+(\Ga^+_\nu\rho^\mu+\rho^\mu\Ga_\nu)\psi.
\end{array}
\eqno(A3) $$ Comparing (A2) with (A3) and considering the
arbitrary $\psi$, we have the algebraic condition to determine the
connection as follows
$$
\Ga^+_\nu\rho^\mu+\rho^\mu\Ga_\nu=\rho^\mu_{;\nu}. \eqno(A4)$$ It
is easy to check
$$
\wt\rho_\mu A \rho^\mu=2{\rm tr}(A),\qquad \wt\rho_\mu \rho^\mu=4,
\eqno(A5) $$ where $A$ is any $2\times2$ matrix. $\wt\rho_\mu$
left times (A4), we have
$$
2{\rm tr}(\Ga_\nu)^+ + 4\Ga_\nu=\wt\rho_\mu\rho^\mu_{:\nu}.
\eqno(A6) $$ Taking trace of (A6), we have
$$
4[{\rm tr}(\Ga_\nu)^+ + {\rm tr}(\Ga_\nu)]={\rm
tr}(\wt\rho_\mu\rho^\mu_{:\nu})=0, \eqno(A7)$$ so we can express
$$
{\rm tr}(\Ga_\nu)=i2 e A_\nu, \eqno(A8)$$ where coefficient $e$
and field $A_\nu$ are real and arbitrary. Then the general
solution of (A6) becomes
$$
\Ga_\mu=\frac 1 4 \wt\rho_\nu\rho^\nu_{;\mu}+ieA_\mu. \eqno(A9)$$
(A9) reflects the consistency of the gravitation with the other
kind interaction. Considering connection $\Ga_\mu$ should just
reflect the geometrical effects, and nonzero $eA_\mu$ results in
physical effects, so we have the geometrical part of the spinor
connection as
$$
\Ga_\mu=\frac 1 4 \wt\rho_\nu\rho^\nu_{;\mu}. \eqno(A10)$$
Similarly we have $\wt\Ga_\mu=\frac 1 4
\rho_\nu\wt\rho^\nu_{;\mu}$.

\begin{thebibliography}{99}
\bibitem{1} B. Saha, {\em Spinor field and accelerated regimes in cosmology},
ArXiv:gr-qc/0512050.
\bibitem{2} M. O. Ribas, F. P. Devecchi, G. M. Kremer, {\em Fermions as sources of accelerated regimes in cosmology},
ArXiv:gr-qc/0511099.
\bibitem{3} B. Saha, T. Boyadjiev,  Phys. Rev. D {\bf 69}, 124010(2004).
\bibitem{4} B. Saha, Phys. Rev. D {\bf 69}, 124006(2004).
\bibitem{gu} Y. Q. Gu, {\em Some properties of the spinor soliton}, Adv. Appl. Cliff. Alg. {\bf 8(1)},
(1997)17,  http://www.clifford-algebras.org/v8/81/gu81.pdf,\\
ArXiv:hep-th/0611210
\bibitem{6} M. Sachs, {\em General relativity and matter} {\bf (Ch.3)}, D. Reidel, 1982.
\bibitem{7} W. L. Bade, H. Jehle, Rev. Mod. Phys.{\bf  25(3)}, (1953)714
\bibitem{8} P. G. Bergmann, Phys. Rev. {\bf 107(2)}, (1957)624
\bibitem{9} D. R. Brill, J. A. Wheeler, Rev. Mod. Phys. {\bf 29(3)}, (1957)465
\bibitem{10} J. P. Crawford, Adv. Appl. Cliff. Alg. {\bf 2(1)}, (1992)75
\bibitem{gu1} Y. Q. Gu, {\em Representation of the Vierbein
Formalism}, ArXiv:gr-qc/0612106.
\bibitem{gu2} Y. Q. Gu, {\em Simplification of the covariant derivatives of spinors}, \\ ArXiv:gr-qc/0610001.


\end{thebibliography}
\end{document}